\documentclass[preprint,floats,prl,aps,groupedaddress]{revtex4}
\usepackage{graphicx}
\usepackage{epstopdf}
\begin{document}
%
\title{Hawking radiation and classical tunneling: a ray phase space approach}
%
\author{\normalsize
  E.R.~Tracy and D. Zhigunov\\ 
   Department of Physics\\
   William \& Mary\\
  Williamsburg, VA 23187-8795, USA\\
}
\noaffiliation
\date{\today}

\begin{abstract}
\noindent Acoustic waves in fluids undergoing the transition from sub- to supersonic flow satisfy governing equations similar to those for light waves in the immediate vicinity of a black hole event horizon. This acoustic analogy has been used by Unruh and others as a conceptual model for `Hawking radiation.' Here we use variational methods, originally introduced by Brizard for the study of linearized MHD, and ray phase space methods, to analyze linearized acoustics in the presence of background flows. The variational formulation endows the evolution equations with natural Hermitian and symplectic structures that prove useful for later analysis. We derive a $2\times 2$ normal form governing the wave evolution in the vicinity of the `event horizon.' This shows that the acoustic model can be reduced locally (in ray phase space) to a standard (scalar) tunneling process weakly coupled to a unidirectional non-dispersive wave (the `incoming wave'). Given the normal form, the Hawking `thermal spectrum' can be derived by invoking standard tunneling theory, but only by ignoring the coupling to the incoming wave. Deriving the normal form requires a novel extension of the modular ray-based theory used previously to study tunneling and mode conversion in plasmas. We also discuss how ray phase space methods can be used to change representation, which brings the problem into a form where the wave functions are less singular than in the usual formulation, a fact that might prove useful in numerical studies.

\end{abstract}

\maketitle

The tilting of light cones in the curved spacetimes of the General Theory of Relativity (GTOR) has an acoustic analog: sound waves propagating in a background with flow. An `event horizon' appears in the acoustic model when the background flow speed matches the local sound speed. This acoustic analogy was first noted by Unruh~\cite{Unruh}, and later examined by Jacobson~\cite{Jacobson}. (See Visser {\em et al.} for a more recent summary of work in this area~\cite{Visser}.) There are other analogs for Hawking radiation that have been explored in the literature, for example Weinfurtner et al.~\cite{Weinfurtner} have experimentally examined surface waves on water flowing in a channel of varying width. In addition to the GTOR literature on this topic (see the references in Visser's article), there has also been some interest in classical analogs of Hawking radiation in the Atomic Molecular and Optical (AMO) literature, where similar effects have been proposed for light wave propagation within certain types of Bose-Einstein condensates. This has led to the study of `artificial black holes'~\cite{ABH}. 

The equations for acoustic wave propagation near the event horizon exhibit an avoided-crossing-type of behavior. (This avoided crossing behavior was noted, for example, by Jacobson~\cite{Jacobson} in the study of black holes using quantum field theory.) Because of the possibility of tunneling between rays, short wavelength fluctuations in the immediate vicinity of the event horizon can escape. While Jacobson cites some of the plasma literature on mode conversion, his method of approach doesn't pursue a ray phase space viewpoint~\cite{Tunneling}. 

The goal of the present paper is to examine the acoustic analogy using ray phase space methods. These methods~\cite{RayTracingBook} can be used to systematically derive the normal form for the evolution equations in the immediate vicinity of the event horizon. This is the simplest possible form for the governing equations, as defined by having the smallest number of terms, arranged in the most symmetrical manner, and the normal form displays those combinations of parameters that are invariant under various transformations, hence it reveals those combinations of parameters that are most important physically. For example, it displays that combination of parameters that will appear in the $\sf S$-matrix connecting the incoming and outgoing wave amplitudes. 

The current problem is not of standard type. It requires an extension of the normal form analysis presented in Appendix F of~\cite{RayTracingBook} in order to identify the uncoupled ray Hamiltonians.  The normal form involves a (scalar) tunneling process weakly coupled to a non-dispersive uni-directional wave (the `incoming wave', also a scalar field). Given the normal form, the Hawking `thermal spectrum' can be derived by invoking standard tunneling theory, but only by ignoring the coupling to the incoming wave. 

The methods presented here can be extended to higher dimensions (for simplicity, in this paper we only discuss the case of one spatial dimension), and to include more complicated physics. For example, the variational methodology used here was originally introduced by Brizard~\cite{Brizard92} to study linearized MHD waves in the presence of flows. But we postpone consideration of those matters because of the greater complexity of the problem if we include magnetic fields. 

The outline of the paper is as follows: In Section~\ref{sec:Heuristics} we present a heuristic treatment of the problem, highlighting key points that might get overshadowed when we dive into the technical details in later sections. In Section~\ref{sec:1DHydro}, we begin by discussing one-dimensional acoustics using a cold fluid model. We then show how to derive the acoustic model (including a background flow) using a variational principle. This will allow us to use standard methods to derive conservation laws, and will lead us to a Hamiltonian formulation of the linearized wave equation. We follow Brizard~\cite{Brizard92} for this part of the paper, and note that the symplectic inner product that plays an important role in our work is also invoked in Weinfurtner~\cite{Weinfurtner}, but they do not use a ray phase space approach. In Section~\ref{sec:WKB}, we examine eikonal solutions of the linearized wave equations and show that we recover the dispersion function~(\ref{eq:DispersionIntro}) away from the event horizon. In the final Section~\ref{sec:NormalForm}, we construct the local wave operator in the vicinity of the horizon using phase space methods to find the normal form. 

Elsewhere, we will discuss the discretization of the phase space variational principle. This provides a means to derive numerical schemes that are symplectic and have good stability properties. 

\section{Heuristic treatment}\label{sec:Heuristics}

Before diving into mathematical details, it is useful to summarize the main result, which is really quite simple if we allow ourselves to ignore some technical matters we will discuss later on. The treatment here is non-relativistic, so transformations between frames are Galilean.

Start with the wave equation for acoustic waves in a {\em uniform} and {\em stationary} background. The dispersion function in that case is the familiar
\begin{equation}
D(k;\omega)\equiv \omega^2 - c_s^2k^2=\left[\omega -c_sk \right]\left[\omega +c_sk \right].
\end{equation}
Here $\omega$ is the wave frequency and $k$ the wavenumber. The two roots of $D=0$ are left- and right-moving nondispersive waves that propagate at the sound speed $\pm c_s$. 

If there is {\em flow} with fluid velocity $v$, Doppler effects must be included, and in the lab frame the dispersion function becomes
\begin{eqnarray}\label{eq:DispersionUniform}\nonumber
D'(k;\omega) &=& (\omega -kv)^2 - c_s^2k^2\\ \nonumber
       &=& \left[\omega -(v+c_s)k \right] \left[\omega -(v-c_s)k \right]\\
                     &\equiv & D_1(k;\omega)D_2(k;\omega).
\end{eqnarray}
Note that if $v=\pm c_s$, a zero-frequency root appears for $D=0.$ This corresponds to a standing (frozen) wave pattern.

  \begin{figure}
    \centering
\includegraphics[width=8cm]{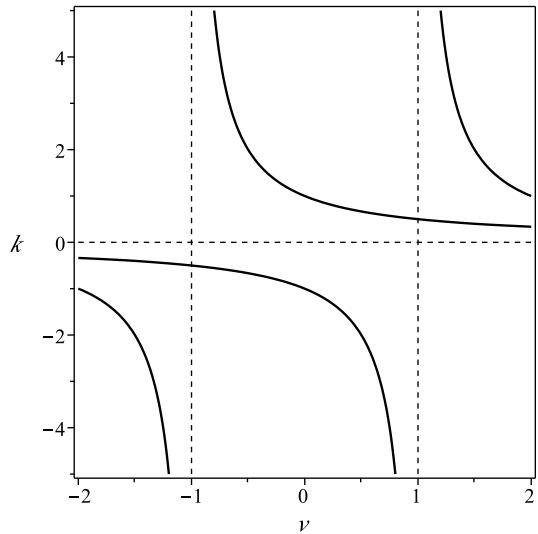}
    \caption{A plot of the surface $D=0$, where $D$ is defined in~(\ref{eq:DispersionUniform}), using $\omega =1$ and $c_{s}=1$. The plot is shown in terms of $k$ and $v$.  To guide the eye, we have also plotted the lines $v=\pm 1$ and $k=0$ (all three of these lines are dashed). The avoided crossings where $v^2=c_{s}^2=1$, are clear.}
    \label{fig:Dispersionkv}
  \end{figure}  

Figure~\ref{fig:Dispersionkv} shows the curves satisfying $D=D_1D_2=0$. These are plotted on the $(v,k)$-plane, rather than ray phase space $(x,k)$, which allows us to see the entire range of behaviors this system can exhibit without having to specify a flow profile. For Figure~\ref{fig:Dispersionkv}, the sound speed is assumed to be constant $c^2_s=1$, implying there are `event horizons' at $v=\pm 1$.

  \begin{figure}
    \centering
\includegraphics[width=8cm]{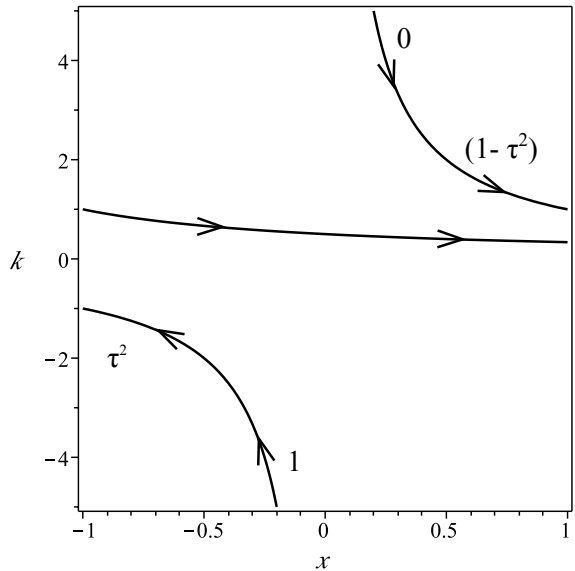}
    \caption{A plot of the surface $D=0$, now restricted to the region around the resonance at $v=+c_s=1$, assuming a linear velocity profile of the form $v(x) = c_s(0)\left(1 +\frac{x}{L} \right)=\left(1 +x \right)$, where we have set $c_s(0)=1$, and $L=1$. Note that this plot uses $x$ as the horizontal axis.  An avoided crossing type of behavior is present at the resonance, where $D_2(x,k)=0$, but there is an additional branch where $D_1(x,k)=0$ is present as well, so this is not tunneling of the standard type. The arrows on the rays indicate the direction of flow of the energy, as given by the Hamilton equations~(\ref{eq:Ham}). The notations $[0,1,\tau^2,(1-\tau^2)]$ are the connection coefficients for the energy on the rays undergoing tunneling, as explained in the text, after Eq.~(\ref{eq:tau}).}
    \label{fig:Dispersion2kv}
  \end{figure}  

We might guess that the local dispersion function governing sound waves in a {\em nonuniform} medium with flow is of the form (we now drop the prime):
\begin{eqnarray}\label{eq:DispersionIntro}\nonumber
D(x,k;\omega) &=& \left[\omega -(v(x)+c_s(x))k \right] 
                                \left[\omega -(v(x)-c_s(x))k \right], \\
            & \equiv & D_1(x,k;\omega)D_2(x,k;\omega).
\end{eqnarray}
The notation we use here (and throughout the paper) emphasizes that, because the background is assumed to be time-stationary, we can treat single-frequency waves by Fourier analysis. 
In these expressions we want to view $D(x,k;\omega)$ as a function of the variables $(x,k)$ which form a conjugate pair on the {\em ray phase space}. The frequency $\omega$ appears as a parameter.

In WKB theory, the dispersion function $D(x,k;\omega)$ is the ray Hamiltonian. In a two-dimensional ray phase space, the zero locus of the Hamiltonian ($D=0$) is also the set of rays. For example, in Figure~(\ref{fig:Dispersion2kv}) we show the rays near the event horizion $v=c_s$ for the special case where $c_s=1$, and a linear velocity profile, $v(x)=1+x$, hence the event horizon is at $x=0$.

The arrows on the rays indicate the direction of the ray evolution given by Hamilton's equations~\cite{RayTracingBook} (subscripts denote partial derivatives):
\begin{equation}\label{eq:Ham}
\frac{d x}{dt} = -D^{-1}_{\omega}D_k, \qquad \frac{d k}{dt} = D^{-1}_{\omega}D_x.
\end{equation}

The smooth central branch that passes near the origin is a right-moving wave associated with the root $D_1=0$. This is the `incoming' wave. A little algebra shows that the group velocity for this wave $\dot x=-D^{-1}_{1\omega}D_{1k}=2c_s$ at the event horizon.

The other branch, associated with $D_2=0$, forms an avoided crossing, typical of tunneling. For those branches of the dispersion surface, as $x$ approaches zero $|k|\rightarrow \infty$. The Hamilton equations for these rays show that disturbances propagate from large- to small-$k$, meaning that to find the disturbances that escape from the vicinity of the event horizon into the subsonic region (to the left in Figure~\ref{fig:Dispersion2kv}) we must specify initial data on the rays {\em at} the event horizon. Upon discretization, we have found the high-$k$ behavior changes dramatically (as will be discussed in a separate paper), hence it would be good to avoid having to specify our initial data in that region, if possible, a point made by Unruh and Jacobson previously.

We should also note that $|k|\rightarrow \infty$ is usually interpreted as the signature of a `resonance.' However, there is nothing fundamental in a mathematical sense about this characterization, since $|k|$ going to infinity at a particular point in $x$ can be removed by a linear canonical transformation. For example, we can use the linear symplectic transformation given in~(\ref{eq:Rotation}), which we will discuss momentarily. A more invariant way to characterize a resonance using normal form theory is given in Chapter 6 of~\cite{RayTracingBook}, and we will use that method here. 

As will be shown in Section~\ref{sec:WKB}, the WKB-based intuition reflected in the expression~(\ref{eq:DispersionIntro}) is valid {\em away} from the event horizon, but it breaks down in the immediate vicinity of the horizon because WKB is invalid there. A local treatment must be developed in order to compute the $\sf S$-matrix connecting incoming and outgoing WKB solutions. We summarize the approach here briefly, with details given in later sections of the paper.

The linearized equations of one-dimensional gas dynamics, Eqs.~(\ref{eq:continuity1})-(\ref{eq:pressure1}), will be derived from a variational principle~(\ref{eq:Variational1}), and a Hamiltonian field theory is then developed using standard methods. The canonical field variables are the particle displacement, $\xi (x,t)$, and the momentum density, $\pi (x,t)$. Written in terms of this canonical formulation, time evolution is governed by a $2\times 2$ Schr\"odinger-type equation~(\ref{eq:SchrodingerForm}). The associated $2\times 2$ Weyl symbol is [see Eqs.~(\ref{eq:SchrodingerEq}) and~(\ref{eq:SymbolD})]:
\begin{equation}\label{eq:ASymbol}
{\sf A}(x,k;\omega) \equiv 
{\sf B}(x,k)-\omega {\sf 1}
=
\left(
\begin{array}{cc}
kv(x)-\omega    &    i \\
-ic_s^2k^2   & kv(x)-\omega
\end{array}
\right).
\end{equation}
To focus on the key points of this heuristic discussion here, we simplify by setting the density $\rho_0$ equal to unity, ignoring Moyal terms in the product $kv(x)$ (defined in~(\ref{eq:Moyal}) below), and using a constant sound speed. 

Notice that the symbol matrix $\sf A$ is {\em not} self-adjoint, but it becomes self-adjoint  (for real $x$, $k$, and $\omega$) when we multiply it by $i$ times the symplectic matrix, $\sf J$, defined in~(\ref{eq:SymplecticJ}). This important fact is a general characteristic of Brizard's theory~\cite{Brizard92}.

To isolate the tunneling process as much as possible, we need to bring the symbol matrix~(\ref{eq:ASymbol}) into normal form. Then the $2\times 2$ operator associated with that symbol will provide us with the simplest {\em local wave equation}, valid in the immediate vicinity of the event horizon. 

To compute the normal form, we start with the eigenvalues and eigenvectors of the {\em non-self-adjoint} $\sf A$. The eigenvalues are:
\begin{equation}\label{eq:DPlusMinus}
D_{\pm}\equiv \left[(v\pm c_s)k-\omega \right],
\end{equation}
with the associated eigenvectors
\begin{equation}\label{eq:Eigenvectors}
{\bf e}_{\pm}\equiv
\frac{1}{(2c_s|k|)^{1/2}}
\left(
\begin{array}{c}
1 \\
\mp ic_s k
\end{array}
\right).
\end{equation}
Note that the $\pm$ sign convention has been chosen to denote the eigenvalue and polarization associated with $v \pm c_s$. The eigenvectors are mutually orthogonal with respect to the complex form of the symplectic inner product, defined as
\begin{equation}\label{eq:SympInner}
\Omega ({\bf u},{\bf w}) \equiv i{\bf u}^{\dag}\cdot {\sf J}\cdot {\bf w}.
\end{equation}
That is $\Omega ({\bf e}_+,{\bf e}_-)=\Omega ({\bf e}_-,{\bf e}_+)=0$. We have normalized them so $\Omega ({\bf e}_{\pm},{\bf e}_{\pm})=\pm 1$.

Let's focus on the event horizon that occurs when $v=c_s$. In that case, the dispersion function $D_-(x,k;\omega)$ is the one associated with the tunneling. For simplicity, let's assume once more that $c_s$ is constant, choose our origin such that this occurs at $x=0$, and linearize $v(x)$:
\begin{equation}
v(x)= c_s\left(1+\frac{x}{L} \right),
\end{equation}
where $L$ is the length scale characteristic of the flow at the event horizon. This implies
\begin{equation}\label{eq:Dminus}
D_-(x,k;\omega)\propto xk - \eta^2, \qquad \eta^2\equiv \frac{L\omega}{c_s}.
\end{equation}
The zero locus in $x$ and $k$ of $D_-(x,k;\omega)=0$, for $\omega \neq 0$, has the characteristic avoided-crossing (hyperbolic) shape for the rays. We will discuss this more in a moment.

Returning to the normal form calculation: for a fixed (but arbitrary) frequency $\omega$, choose $k_0(\omega)$ by solving $D_+(x=0,k_0(\omega);\omega)=0$:
\begin{equation}
D_+=0 \qquad \Rightarrow \qquad k_0(\omega) = \frac{\omega}{2c_s}.
\end{equation}
Now construct the {\em constant} (in $x$ and $k$) matrix $\sf Q(\omega)$ from the eigenvectors:
\begin{equation}
{\sf Q}(\omega) \equiv \frac{1}{\left(2c_sk_0 \right)^{1/2}}
\left(
\begin{array}{cc}
1                             &           1 \\
-ic_sk_0(\omega)   & +ic_sk_0(\omega)
\end{array}
\right).
\end{equation}
Note that ${\sf Q}(\omega)$ is not unitary, but instead satisfies
\begin{equation}\label{eq:QdagJQ}
i{\sf Q}^{\dag}(\omega)\cdot {\sf J} \cdot {\sf Q}(\omega) =
\left(
\begin{array}{cc}
1  &  0  \\
0  & -1
\end{array}
\right).
\end{equation}
Note also that ${\sf Q}(\omega)$ is non-invertible when $\omega \rightarrow 0$. This is a singular limit, reflected in the fact that~(\ref{eq:QdagJQ}) remains true for all $\omega \neq 0$.

The next step is to compute the new symbol matrix
\begin{equation}
{\sf D}(x,k;\omega) \equiv i{\sf Q}^{\dag}(\omega)\cdot {\sf J} \cdot 
                                                   {\sf A}(x,k;\omega)\cdot {\sf Q}(\omega).
\end{equation}
This new matrix is self-adjoint by construction because $i{\sf J} \cdot {\sf A}$ is self-adjoint for all real $x$, $k$ and $\omega$. It is important to note the use of a {\em congruence} transformation here, not a similarity transformation. The reason that a congruence appears here is that we derive the wave operator [in Section~\ref{sec:1DHydro}, see Eq.~(\ref{eq:Action1})] from a variational principle, which for the linearized dynamics is a bi-linear form. More details regarding the appearance of congruence transformations in ray phase space methods for multicomponent wave equations can be found in Appendix C.2 of~\cite{RayTracingBook}.

We should note that {\em diagonalizing} ${\sf D}(x,k;\omega)$, as opposed to bringing it into normal form, requires use of the eigenvectors of $\sf D$, not $\sf A$. The eigenvectors of ${\sf D}(x,k;\omega)$ are $x$- and $k$-dependent. This leads to problems when we try to apply the congruence transformation within the variational principle~(\ref{eq:Action1}), because the entries of the congruence transformation $\sf Q$ become operators, not constants~\cite{Friedland1987}. In addition, the eigenvalues of ${\sf D}(x,k;\omega)$ are complicated functions of $x$, $k$, and $\omega$, because they include the coupling. The coupling mixes the different wave behaviors and makes physical interpretation more difficult. The normal form, in contrast, isolates as much as possible the {\em uncoupled} dispersion functions, closely related to~(\ref{eq:DPlusMinus}).

The next step in the normal form calculation is to Taylor expand ${\sf D}(x,k;\omega)$ about $k=k_0(\omega)$, writing $k=k_0+\kappa$. We keep only the leading terms in $\kappa$ in each of the four entries. (Higher order corrections can, of course be included if they are deemed necessary. See~\cite{HigherOrder} for a discussion.) After a little algebra, we find
\begin{eqnarray}\label{eq:NormalForm}\nonumber
{\sf D}(x,k;\omega) &\equiv&
\left(
\begin{array}{cc}
D_a    &    \alpha \\
\alpha     &     D_b
\end{array}
\right) \\
&\approx& 
\left(
\begin{array}{cc}
(v+c_s)\kappa-\omega    &    -ic_s\kappa \\
ic_s\kappa   & (c_s-v)\kappa+\omega
\end{array}
\right).
\end{eqnarray}
We identify $D_a$ and $D_b$ as the uncoupled dispersion functions at the event horizon, and $\alpha$ is the coupling between them. (Note that the eigenvalues of ${\sf A}(x,k;\omega)$ given in~(\ref{eq:DPlusMinus}), $D_+$ and $D_-$, depend upon $k$, while $D_a$ and $D_b$ depend upon $\kappa$.)

The coupling is usually evaluated at the base point of the Taylor expansion in ray phase space [here $(x=0,k=k_0)$], setting it equal to a constant value~\cite{RayTracingBook}. This simplification is often sufficient to get good results. But in the present case, setting $\kappa = 0$ implies zero coupling. We carry the coupling term along to remind ourselves it is not exactly zero. Neglect of the coupling is required to recover the Hawking result outlined below.

The entry $D_a$ is the dispersion function for a right-moving wave with group velocity $2c_s$ at the event horizon, as can easily be verified using the Hamilton equations with $D_a$ as ray Hamiltonian. This is the `incoming wave'.

The entry $D_b$ is the wave undergoing tunneling when $v\approx c_s$. We now consider this avoided crossing separately.  
\begin{eqnarray}\label{eq:Db}\nonumber
D_b(x,\kappa;\omega) &\equiv& (v-c_s)\kappa - \omega \\ \nonumber
                                     &=& \frac{c_s}{L}\kappa x -\omega \\
                                     &=& \frac{c_s}{L}\left(x\kappa -\eta^2 \right).
\end{eqnarray}
We emphasize once more that while $D_-(x,k;\omega)$ in~(\ref{eq:Dminus}) is an eigenvalue of ${\sf A}(x,k;\omega)$, $D_b(x,\kappa;\omega)$ is a diagonal element of the normal form~(\ref{eq:NormalForm}), and a function of $\kappa = k-k_0(\omega)$, not $k$. 

The ray equations, using $D_b$ as the ray Hamiltonian, are:
\begin{equation}\label{eq:Ham2}
\frac{d x}{dt} = -{D}^{-1}_{b,\omega}{D}_{b,\kappa} =\frac{c_s}{L} x, \qquad \frac{d \kappa}{dt} = {D}^{-1}_{b,\omega}{D}_{b,x} = -\frac{c_s}{L} \kappa.
\end{equation}
This implies that disturbances that start near $x\approx 0$ at very small spatial scales (large $|k|$) propagate toward smaller $|k|$, while moving away from the origin in $x$. In particular, the ray starting with $x\approx 0$ and large {\em negative} $k$ leaves the region of the resonance traveling to the left, i.e. it `escapes' from the event horizon and propagates into the subsonic region $x<0$.

Now consider the following linear canonical transformation:
\begin{equation}\label{eq:Rotation}
x = \frac{1}{\surd 2}\left(X+K \right), \qquad \kappa = \frac{1}{\surd 2}\left(X-K \right).
\end{equation}
This is a rotation by $45^{\circ}$, and it puts~(\ref{eq:Db}) into the tunneling normal form (see, for example, page 243, Eq.~(6.41), and Figure 6.7 on page 244 of~\cite{RayTracingBook}):
\begin{equation}\label{eq:ParabolicCylinder}
D_b'(X,K;\omega) \equiv \frac{c_s}{L}\left[\frac{1}{2} \left(X^2 - K^2\right)-\eta^2(\omega)\right].
\end{equation}

Linear canonical transformations on ray phase space induce related unitary transformations in the associated Hilbert space of wave functions. This change of representation is a generalization of the Fourier transform, called the {\em metaplectic} transform. For example, the unitary transform that takes wave functions in the $x$-representation to wave functions in the $X$-representation is:
\begin{equation}\label{eq:MetaplecticxX}
 \phi (X) = \frac{1}{\sqrt{2\pi}}\int dx \; e^{-iF_1(X,x)} \psi (x),
\end{equation}
where
\begin{equation}
F_1(X,x) \equiv \frac{1}{2}(x^2-\surd 2 xX +X^2).
\end{equation}
The inverse of~(\ref{eq:MetaplecticxX}) is simply
\begin{equation}\label{eq:MetaplecticxXInv}
 \psi (x) = \frac{1}{\sqrt{2\pi}}\int dx \; e^{iF_1(X,x)} \phi (X),
\end{equation}
See Appendix E of~\cite{RayTracingBook} for details.

Because $X$ and $K$ form a canonical pair, if we choose the $X$-representation to write our wave equation, we have the familiar association
\begin{equation}
K \leftrightarrow -i\frac{\partial}{\partial X},
\end{equation}
implying that the equation governing the mode shape for frequency $\omega$ is:
\begin{equation} \label{eq:NormalFormX}
\left[X^2 -2\eta^2(\omega)\right]\psi(X)+\frac{\partial ^2\psi}{\partial X^2}=0.
\end{equation}
The solution of this equation can be written in terms of parabolic cylinder functions, and the $\sf S$-matrix elements connecting incoming and outgoing rays computed. The transmission coefficient (relating the outgoing and incoming {\em amplitude} on the ray) is
\begin{equation}\label{eq:tau}
\tau \equiv e^{-\pi \eta^2} = e^{-\pi \omega L/c_{s}}.
\end{equation}
(Note that in the case of tunneling the incoming ray connects smoothly to the {\em transmitted} ray, not the converted ray. See Section 6.2 and Problems 6.4 and 6.5 of~\cite{RayTracingBook} for details.)

As already mentioned, the linear canonical transformation~(\ref{eq:Rotation}) from $(x,\kappa)\rightarrow (X,K)$ generates a unitary transformation on the Hilbert space of wave functions. Therefore, so long as we keep track of incoming and outgoing pairings (using the rays), the $\sf S$-matrix elements are unchanged. This allows us to compute the connection coefficients for the original $x$-representation 

The {\em energy} associated with the outgoing transmitted ray is proportional to the square of the wave amplitude, so it is related to the energy on the incoming ray using $\tau^2$. Total energy is conserved, so the energy on the {\em converted} ray is proportional to $(1-\tau^2)$. In Figure~\ref{fig:Dispersion2kv} we show, for example, the energies for the transmitted and converted rays for an initial disturbance that starts with unit amplitude and large {\em negative} $k$.

To this point, these results are entirely classical. The connection with `Hawking radiation' from a black hole is through the following analogy. Rewrite the exponent in the {\em energy} transmission coefficient as [from~(\ref{eq:tau}) we find $\tau^2=\exp(-2\pi \eta^2)$]:
\begin{equation}
\frac{2\pi \omega L}{c_{s}} \equiv \frac{\hbar \omega}{k_B T_{eff}},
\end{equation}
where $\hbar$ is Planck's constant, $k_B$ is Boltzmann's constant, and $T_{eff}$ is an `effective temperature':
\begin{equation}\label{eq:Teff}
T_{eff} \equiv \frac{\hbar c_{s}}{2\pi k_B L}.
\end{equation}
Note that this effective temperature is inversely proportional to the length scale. (In the famous result by Hawking, the effective temperature of a black hole is inversely proprtional to the mass, while the Schwarzschild radius is proportional to the mass. This means that the temperature and characteristic length scale at the event horizon are in the same inverse relation as here.) 

If we consider laboratory acoustics and transitional flow, for example with jet nozzles, we can choose approximate values for the transition region at the throat of the nozzle. For example, following Unruh, we choose the length scale $L\approx 10^{-3}\; m$, and the sound speed, $c_s\approx 300 \; m/s$. This gives
\begin{equation}
\tau (\omega) \approx \exp (-10^{-6}\omega),
\end{equation}
implying that we can only observe the `thermal' character of the emission if we look in the MHz range of frequencies, and this in an extremely turbulent environment. (Unruh~\cite{Unruh} acknowledges that this is a challenging measurement to make, though he points out that it is easier than using a laboratory-scale black hole!)

Our primary interest in this paper is in the theoretical formalism, in particular  an examination of the Unruh model from the perspective of ray phase space. The normal form method presented here requires a modification of earlier methods, and the symplectic inner product~(\ref{eq:SympInner}) plays a key role, which is new. The extension used here should be applicable to other problems where this symplectic structure appears (e.g. all those covered by Brizard's theory of linearized MHD). 

This completes our summary of the main points. We now move to the derivation of the acoustic model from one-dimensional gas dynamics, the canonical formulation, and WKB.

\section{One-dimensional hydrodynamics}\label{sec:1DHydro}

The approach followed here is a special case of the general theory presented in Brizard~\cite{Brizard92}. We use a cold ideal fluid model in one spatial dimension, $x$, which has as dependent variables the density, $\rho (x,t)$, the velocity, $v(x,t)$, and the pressure $p(x,t)$. These quantities obey the following evolution equations:
\begin{equation}\label{eq:MassCon}
\rho_t + \left(\rho v \right)_x = 0,
\end{equation}
\begin{equation}\label{eq:MomentumCon}
\rho \left(v_t + vv_x \right) = -p_x,
\end{equation}
and
\begin{equation}\label{eq:Adiabatic}
p_t + vp_x = -\gamma pv_x,
\end{equation}
where the subscripts denote partial derivatives and $\gamma$ is the ratio of specific heats. We recognize the first equation as the statement of mass conservation, the second of momentum conservation, and the third is required for the evolution following fluid trajectories to be adiabatic with the equation of state $p \propto \rho^{\gamma}$. 

\subsection{Acoustic waves}

It is important to note that the derivation of the acoustic model given below is non-physical in the sense that we start with nonlinear one-dimensional inviscid fluid flow, then linearize around a given (time stationary) spatial profile in density and background flow. This means that we are ignoring the formation of shocks, which are well-known to occur in this system of PDEs.

Now linearize Eqs.~(\ref{eq:MassCon}), (\ref{eq:MomentumCon}), and~(\ref{eq:Adiabatic}) to find the evolution equation for small amplitude (acoustic) waves in a nonuniform, stationary, background with flow. Write $\rho (x,t)=\rho_0(x) +\epsilon \rho_1(x,t)$, $v_0(x,t)=v_0(x)+\epsilon v_1(x,t)$, and $p(x,t) =p_0(x) +\epsilon p_1(x,t)$. The constant $\epsilon$ is a formal expansion parameter. 
Expand in powers of $\epsilon$ and collect terms. The zeroth order equilibrium must satisfy
\begin{eqnarray}\label{eq:density0}
(\rho_0v_0)_x &=& 0, \\ \label{eq:energy0}
\left(\frac{1}{2}\rho_0v_0^2 +p_0\right)_x &=& 0, \\
v_0p_{0x} &=& -\gamma p_0v_{0x}.
\end{eqnarray}
We note that the last condition, combined with the first, implies $p_0 \propto  \rho_0^{\gamma}$. Notice, further, that these equilibrium conditions imply that the pressure, density, and velocity profiles are not independent. In particular, the sound speed profile is [see Eq.~(\ref{eq:SoundSpeed2})]:
\begin{equation}\label{eq:SoundSpeed1}
c_s^2(x) = \left.\frac{dp_0}{d\rho_0}\right|_x = \gamma \frac{p_0(x)}{\rho_0(x)}.
\end{equation}
This is not independent of the background velocity profile, $v_0(x)$. A little algebra shows that
\begin{equation}
\frac{c^2_s(x)}{c^2_s(0)} =\left(\frac{v_0(0)}{v_0(x)} \right)^{\gamma -1}.
\end{equation}
This, of course, does not preclude these two velocities from becoming equal one another, which is the situation of interest to us.

At first order in $\epsilon$ we have
\begin{eqnarray}\label{eq:continuity1}
\rho_{1t} &=&-\left(\rho_0v_1+\rho_1v_0\right)_x,\\
\label{eq:momentum1}
\rho_0 v_{1t}  &=&-\frac{1}{2}\rho_1\left(v_0^2 \right)_x-\rho_0\left(v_0v_1\right)_x -p_{1x},\\ \label{eq:pressure1}
p_{1t} &=& -\left(v_0p_{1x} +\gamma p_1v_{0x}\right) - \left(v_1p_{0x} +\gamma p_0v_{1x}\right).
\end{eqnarray}
(N.B. The first term on the RHS of Eq.~(\ref{eq:momentum1}) is missing in Eq.(15) of~\cite{Brizard92}.) Following Brizard, we now replace the three fields $(\rho_1,v_1,p_1)$ with the first order particle displacement $\xi$. The variations in the field quantities are determined by the particle motions through the following identities
\begin{eqnarray}\label{eq:densityxi}
\rho_{1} &=&-\left(\rho_0\xi\right)_x,\\
\label{eq:velocityxi}
v_{1}  &=& \xi_t +v_0\xi_x -\xi v_{0x},\\
p_{1} &=& -\xi p_{0x} -\gamma p_0\xi_x.
\end{eqnarray}

These identities are inserted into the first-order evolution equations to derive the evolution equation for the particle displacement. In particular, the first-order momentum conservation law~(\ref{eq:momentum1}) becomes, after some straightforward but lengthy algebra, 
\begin{eqnarray}\label{eq:xievolution}
\rho_0 \xi_{tt} +2\rho_0v \xi_{xt} &=& \frac{\partial }{\partial x}\left(\gamma p_0-\rho_0v_0^2\right)\frac{\partial \xi}{\partial x},\\ \nonumber
						     &\equiv& {\widehat F}(\xi).
\end{eqnarray}
Note that this result requires use of the zeroth order equilibrium conditions~(\ref{eq:density0}) and~(\ref{eq:energy0}). Note also that~(\ref{eq:xievolution}) agrees with Eq.~(22) of Brizard~\cite{Brizard92}, when that expression is reduced to one spatial dimension, and zero magnetic field.

Now introduce the following variational principle (the overall factor of $1/2$ will ensure that the canonical momentum is equal to the physical momentum density.)
\begin{equation}\label{eq:Variational1}
{\cal A}[\xi] \equiv \frac{1}{2}\int_{t_0}^{t_1} dt \left\{\int_{-\infty}^{+\infty}\; dx \left[ \rho_0(\xi_t+v\xi_x)^2-\gamma p_0\xi_x^2\right]\right\},
\end{equation}
A standard calculation shows that the evolution equation~(\ref{eq:xievolution}) follows when we require stationarity with respect to the variation $\delta \xi (x,t)$, assuming the stationary background obeys~(\ref{eq:density0}). From this point, unless otherwise noted, time integrals are from $t_0$ to $t_1>t_0$, and all spatial integrals from $-\infty$ to $+\infty$.

The variational principle~(\ref{eq:Variational1}) can be used to construct a Hamiltonian formulation using the following standard algorithm. First, define the Lagrangian density
\begin{equation}
{\cal L} \equiv \frac{1}{2} \rho_0(\xi_t+v_0\xi_x)^2 + \frac{1}{2}\xi {\widehat G}\xi .
\end{equation}
where
\begin{eqnarray}\label{eq:Gdef}\nonumber
{\widehat G}\xi\ &\equiv& {\widehat F}\xi +\frac{\partial }{\partial x} \rho_0v_0^2 \frac{\partial \xi}{\partial x},\\
                  & = & \gamma \frac{\partial }{\partial x} p_0 \frac{\partial \xi}{\partial x}
\end{eqnarray}

The canonical momentum density is
\begin{eqnarray}\nonumber
\pi (x,t) &\equiv& \frac{\partial {\cal L}}{\partial \xi_t},\\
             &   =  &   \rho_0\left( \xi_t+v_0\xi_x\right),
\end{eqnarray}
which we identify as the physical momentum density. The Hamiltonian density is constructed using the Legendre transformation (first writing $\xi_t = \rho_0^{-1}\pi-v_0\xi_x$)
\begin{eqnarray}
{\cal H}    &\equiv&  \pi \xi_t (\xi, \pi)- {\cal L}(\xi, \pi), \\
                 &     =    & \frac{1}{2}\frac{\pi^2}{\rho_0} -v_0\pi \xi_x  -\frac{1}{2} \xi {\widehat G}\xi.
\end{eqnarray}
Notice that this leads to a Hamiltonian density that is quadratic in $\xi$ (through its derivatives) and $\pi$, as expected. The Hamiltonian itself is the integration of the density over $x$ and $t$, which we write in the form:
\begin{equation}
{\cal H}[\xi, \pi] \equiv  \int dt \; \left[ \frac{1}{2}\rho_0^{-1}\langle \pi | \pi \rangle -\langle \pi |v_0\xi_x \rangle-\frac{1}{2}\langle \xi | {\widehat G}\xi \rangle \right],
\end{equation}
where we have introduced the inner product notation
\begin{equation}\label{eq:InnerProduct}
\langle \chi |\lambda \rangle \equiv \int dx \;\chi^*(x)\lambda(x).
\end{equation}

A little algebra verifies that
\begin{eqnarray}\label{eq:HamsEqXi}
\xi_t &\equiv& \frac{\delta {\cal H}}{\delta \pi} = \rho_0^{-1} \pi - v_0\xi_x, \\ \label{eq:HamsEqPi}
\pi_t &\equiv& -\frac{\delta {\cal H}}{\delta \xi} = -(v_0\pi)_x +{\widehat G}\xi.
\end{eqnarray}
The first of this pair of evolution equations is simply a rewriting of the relationship between $\xi_t$ and $\pi$, while the second is seen to be a recasting of~(\ref{eq:xievolution}), after using~(\ref{eq:Gdef}).

The pair of canonical evolution equations~(\ref{eq:HamsEqXi}) and~(\ref{eq:HamsEqPi}) can be shown to have a Hermitian structure. This will prove valuable in deriving the normal form. First, introduce the $2\times 2$ symplectic matrix 
\begin{equation}\label{eq:SymplecticJ}
{\sf J} \equiv
\left(
\begin{array}{cc}
 0 &  1 \\
 -1 & 0
\end{array}
\right).
\end{equation}
Second, define the two-component {\em complex} field $\psi\equiv (\xi,\pi)^T$. Third, the {\em complex} symplectic product of $\psi_a$ and $\psi_b$ is now introduced:
\begin{eqnarray}
\omega (\psi_a,\psi_b) &\equiv& \psi_a^{\dag}\cdot {\sf J} \cdot \psi_b,\\
                                     &=& \xi_a^*\pi_b - \pi_a^* \xi_b.
\end{eqnarray}
(The values of $x$ and $t$ are identical, the symplectic product concerns the two-component vector indices.) Finally, define the (degenerate) canonical inner product on the linear symplectic complex vector space
\begin{eqnarray}\label{eq:CanInner}\nonumber
\langle \psi_a|\psi_b \rangle_{can} &\equiv& i \int dx (\xi^*_a\pi_b -\pi^*_a\xi_b), \\
                                                       &=& \langle \psi_b|\psi_a \rangle_{can}^*
\end{eqnarray}

The evolution equations~(\ref{eq:HamsEqXi}) and~(\ref{eq:HamsEqPi}) are now written in the Schr\"odinger form
\begin{equation}\label{eq:SchrodingerForm}
i \frac{\partial \psi^{\alpha}}{\partial t} \equiv {\widehat {\cal B}}^{\alpha}_{\gamma}\psi^{\gamma}.
\end{equation}
where we sum over repeated indices. Explicitly:
\begin{equation}\label{eq:SchrodingerEq}
i\frac{\partial}{\partial t}
\left(
\begin{array}{c}
\xi \\
\pi
\end{array}
\right)
=
\left(
\begin{array}{cc}
-iv_0\partial_x &  i\rho_0^{-1}\\
i{\widehat G}          &   -i\partial_xv_0
\end{array}
\right)
\left(
\begin{array}{c}
\xi \\
\pi
\end{array}
\right).
\end{equation}
Unless otherwise noted, derivatives act on all quantities {\em to the right}. 

The Weyl symbol of an operator ${\widehat {\cal A}}$ is defined as~\cite{RayTracingBook}:
\begin{equation}
a(x,k) \equiv \int ds e^{-iks}\langle x+\frac{s}{2}\left| {\widehat {\cal A}}\right|x-\frac{s}{2}\rangle.
\end{equation}
The Weyl symbol mapping, which is invertible, takes operators and maps them to functions on ray phase space:
\begin{equation}
{\widehat {\cal A}} \quad \stackrel{\Sigma}{\longleftrightarrow} \quad a(x,k),
\end{equation}
This mapping is linear, and topological, meaning that it preserves neighborhood relations in the two spaces. Since operators generally don't commute, this implies that the symbol of the operator product ${\cal A}_1{\cal A} _2$ cannot be simply the product of the related symbols $a_1(x,k)$ and $a_2(x,k)$. In fact the symbol of the product is given by the {\em Moyal} product, denoted
\begin{equation}
{\widehat {\cal A}}_1{\widehat {\cal A}}_2 \quad \stackrel{\Sigma}{\longleftrightarrow} \quad a_1(x,k)*a_2(x,k),
\end{equation}
where
\begin{eqnarray}\label{eq:Moyal}\nonumber
a_1(x,k)*a_2(x,k) \equiv \sum_{n=0}^{\infty} \left(\frac{i}{2}\right)^n \frac{1}{n!} &\times &  \\
 a_1(x,k)
  \left(\stackrel{\leftarrow}{\partial_x}
          \stackrel{\rightarrow}{\partial_k} -
           \stackrel{\leftarrow}{\partial_k}
          \stackrel{\rightarrow}{\partial_x}
           \right)^n
a_2(x,k). &&
\end{eqnarray}
The reader is referred to Chapter 2 of~\cite{RayTracingBook} for a complete discussion. Here we simply quote results unproven.

Using~(\ref{eq:Moyal}), the Weyl symbol of ${\widehat {\cal B}}$ is 
\begin{equation}\label{eq:SymbolD}
{\sf B}(x,k)
\equiv
\left(
\begin{array}{cc}
v_0(x)*k &  i\rho^{-1}_0(x)\\
iG(x,k)          &   k*v_0(x)
\end{array}
\right).
\end{equation}
From~(\ref{eq:Gdef}) we have 
\begin{equation}\label{eq:Gsymbol}
G(x,k) \equiv -\gamma k*p_0(x)*k.
\end{equation}
Note that, although the symbol matrix $\sf B$ is not self-adjoint, the symbol matrix
\begin{equation}
{\sf D}(x,k)\equiv i{\sf J}\cdot {\sf B}(x,k) =
\left(
\begin{array}{cc}
\gamma k*p_0(x)*k          &   ik*v_0(x) \\
-iv_0(x)*k &  \rho^{-1}_0(x)
\end{array}
\right),
\end{equation}
{\em is} self-adjoint. The symbol $k*p_0(x)*k$ is real, and the symbol $k*v_0(x)$ is the complex conjugate of the symbol $v_0(x)*k$, for real $x$ and $k$. These facts follow from the properties of the Moyal product~(\ref{eq:Moyal}).

Given an inner product (in this case the canonical inner product), the adjoint of any operator $\widehat{\cal O}$ is that unique operator ${\widehat{\cal O}}^{\dag}$ defined by the property
\begin{equation}
\langle {\widehat{\cal O}}^{\dag}\psi_a|\psi_b\rangle_{can} \equiv \langle \psi_a| {\widehat{\cal O}}\psi_b\rangle_{can} ,
\qquad \forall \psi_a,\psi_b.
\end{equation}

Therefore, a little algebra shows that the adjoint evolution equation is
\begin{equation}
-i \frac{\partial \psi^{\dag \alpha}}{\partial t} \equiv \psi^{\dag \gamma}{\widehat {\cal B}}^{\dag \alpha}_{\gamma}.
\end{equation}
where
\begin{equation}
{\widehat {\cal B}}^{\dag} \equiv
\left(
\begin{array}{cc}
i\stackrel{\leftarrow}{\partial_x}v & -i\stackrel{\leftarrow}{\widehat G}   \\
-i\rho_0^{-1}       &   iv\stackrel{\leftarrow}{\partial_x}
\end{array}
\right).
\end{equation}

In the current case, it is possible to show that ${\widehat {\cal B}}$ is self-adjoint with respect to the {\em canonical} inner product $\langle \;| \;\rangle_{can}$:
\begin{eqnarray}\nonumber
\langle \psi_a |{\widehat {\cal B}}\psi_b \rangle_{can} &=& 
                   i\int dx \; \left(\xi_a^* ,\; \pi_a^* \right)\cdot {\sf J} \cdot {\widehat {\cal B}} \cdot 
                                 \left(
							\begin{array}{c}
							\xi_b\\
							\pi_b
							\end{array}
							\right), \\ \nonumber
    					         &=& \int dx \; 
	\left[\gamma p_0\xi_{ax}^*\xi_{bx} + \xi_a^*(v_0\pi_b)_x \right. \\
						  & & \qquad \quad\left. -v_0 \pi_a^*\xi_{bx}+\rho_0^{-1}\pi_a^*\pi_b \right]
\end{eqnarray}
After integration by parts, a little algebra shows that
\begin{equation}
\langle \psi_a |{\widehat {\cal B}}\psi_b \rangle_{can} = 
              \langle {\widehat {\cal B}}\psi_a |\psi_b \rangle_{can}, \qquad \forall \psi_a,\psi_b,
\end{equation}
implying that ${\widehat {\cal B}}^{\dag}={\widehat {\cal B}}$ with respect to $\langle \; |\; \rangle_{can}$ as claimed.

Before discussing the WKB analysis of the evolution equations, we note that using the concepts we now have in hand, we can introduce the {\em phase space} variational principle:
\begin{eqnarray}\label{eq:Action1}\nonumber
{\cal A}'[\psi] &\equiv& i\int dt  \langle {\psi},(i\partial_t-{\widehat {\cal B}})\psi \rangle_{can} ,\\
					&=& \int dt  \left[\langle {\psi}_t,\psi\rangle_{can}-i\langle {\psi},{\widehat {\cal B}}\psi \rangle_{can} \right].
\end{eqnarray}
Explicitly:
\begin{eqnarray}\label{eq:Variation1}\nonumber
&{\cal A}'[\psi]& = i\int dt \; dx {\Bigg \{}(\xi^*_t\pi -\pi^*_t\xi ) -  \\ 
          &&  \left[ \gamma p_0|\xi_{x}|^2 + \xi^*(v_0\pi)_x -v_0 \pi^*\xi_{x}+\frac{|\pi|^2}{\rho_0} \right]\Bigg \}
\end{eqnarray}
This variational principle will prove useful when we derive the $2\times 2$ normal form. Also, it is the starting point to discretize the dynamics and derive symplectic integrators, which we will discuss in a separate paper. 

It is sometimes useful to write the phase space variational principle in terms of {\em real} canonical fields $(\xi, \pi)$, this variational principle becomes
\begin{eqnarray}\label{eq:Action2}\nonumber
{\bar {\cal A}}[\xi, \pi] &\equiv& \int dt \; dx \Bigg \{\frac{1}{2}(\xi_t\pi -\pi_t\xi )- \\ & & \left(\frac{\gamma p_0}{2}\xi^2_{x} 
											- \pi v_0\xi_x +\frac{\pi^2}{2\rho_0} \right)\Bigg \}.
\end{eqnarray}
We have integrated by parts, and introduced an overall constant factor of $i/2$ in order to cast this into a more standard form, using the fact that overall factors in the variational principle do not affect the resulting evolution equations. A short calculation leads back to Hamilton's equations~(\ref{eq:HamsEqXi}) and (\ref{eq:HamsEqPi}), as required. 

\section{WKB analysis}\label{sec:WKB}

Now we introduce a single-frequency eikonal ansatz: 
\begin{equation}\label{eq:EikonalAnsatz}
\psi (x,t) \equiv
a(x)e^{i\left[\theta(x)-\omega t\right]}
\left(
\begin{array}{c}
e_{\xi}(x)\\
e_{\pi}(x)
\end{array}
\right),
\end{equation}
where the `polarization' ${\hat {\bf e}}(x)\equiv (e_{\xi},e_{\pi})^T$ is assumed to vary on the same length scale as the background. Use this ansatz in~(\ref{eq:HamsEqXi}) and~(\ref{eq:HamsEqPi}). At leading order, assuming the derivative acts only on the phase, we get:
\begin{equation}\label{eq:WaveEq2}
{\sf A}(x,k;\omega)\cdot {\hat {\bf e}}(x) \equiv
\left(
\begin{array}{cc}
v_0k(x)-\omega               & i\rho_0^{-1}    \\
iG(x,k)    & v_0k(x)-\omega   
\end{array}
\right)\left(
\begin{array}{c}
e_{\xi} \\
e_{\pi}
\end{array}
\right)=0,
\end{equation}
where
\begin{equation}
k(x) \equiv \frac{d\theta}{dx},
\end{equation}
and
\begin{equation}
G(x,k) \equiv -\gamma p_0(x) k^2(x) = -\rho_0c^2_{s}k^2(x).
\end{equation}
We have introduced the sound speed (invoking the equation of state $p_0 \propto \rho_0^{\gamma}$): 
\begin{equation}\label{eq:SoundSpeed2}
c_s^2 = \frac{dp_0}{d\rho_0} = \gamma \frac{p_0}{\rho_0}.
\end{equation}

For there to be non-trivial solutions to~(\ref{eq:WaveEq2}), for a given $\omega$ and $x$, at least one of the eigenvalues of ${\sf A}[x,k=\theta_x(x);\omega]$ must vanish. These eigenvalues are denoted
\begin{equation}
D_{\pm}[x,k=\theta_x(x);\omega]\equiv \omega - \left[v_0(x)\pm c_s(x) \right]k(x).
\end{equation}
In a slight abuse of notation, we have retained $D_{\pm}$ to denote the ray Hamiltonians, even though they are eigenvalues of $\sf A$, not $\sf D$.
These are a pair of Hamiltonians, one for each of the two types of {\em uncoupled} rays. Rays of each type live on the surfaces $D_{\pm}(x,k;\omega)=0$. 

The associated Hamilton's equations are
\begin{eqnarray}\label{eq:HamsEqs}
\left. \frac{dx}{dt}\right|_{\pm} &=& v_0(x)\pm c_s(x), \\
\left. \frac{dk}{dt}\right|_{\pm} &=& -\left[v_0(x)\pm c_{s}(x) \right]_xk.
\end{eqnarray}

We can find the associated polarization for each ray by assuming the relevant dispersion relation is satisfied, using it in~(\ref{eq:WaveEq2}), and then solving for the associated null eigenvector. For example, write $\omega =(v\pm c_s)k $. The null eigenvectors must satisfy
\begin{equation}\label{eq:WaveEq3}
\left(
\begin{array}{cc}
\mp c_sk                & i\rho_0^{-1}    \\
-i \rho_0c^2_{s}k^2    & \mp c_sk  
\end{array}
\right)\left(
\begin{array}{c}
e_1 \\
e_2
\end{array}
\right)_{\pm}=0.
\end{equation}
The associated polarization vectors are
\begin{equation}\label{eq:polarization}
{\hat {\bf e}}_{\pm}(x,k)=\frac{1}{\left(2\rho_0 c_sk \right)^{1/2}}\left(
\begin{array}{c}
1 \\
\mp \rho_0c_sk
\end{array}
\right),
\end{equation}
where $\rho_0$, $c_s$, are functions of $x$, and $k$ is treated as a free parameter here. These have been normalized with respect to the symplectic inner product, as per the discussion after~(\ref{eq:SympInner}), which simplifies later expressions. To use this result in the eikonal solution, we set $k(x)=d\theta/dx$, {\em after} we have solved for $\theta(x)$, as described below. 

The amplitude $a(x)$ varies in a manner governed by the action conservation law, which we can derive as follows. Choose one of the two polarizations, for example the `$-$' polarization. Using the polarization~(\ref{eq:polarization}) with $k(x)$ unspecified as yet, insert the ansatz~(\ref{eq:EikonalAnsatz}) into the variational principle~(\ref{eq:Variation1}), keeping only the leading order terms. I.e. the derivatives act only $\theta(x)$, but not the background quantities, or $a(x)$ and $k(x)$, consistent with the leading order eikonal approximation. After some algebra, this gives:
\begin{equation}\label{eq:Variation3}\nonumber
{\bar {\cal A}}[a,\theta] = \int \; dx \; \left[\omega - (v_0-c_s)\theta_x\right]a^2 (x).
\end{equation}

Stationarity with respect to variations in the amplitude implies 
\begin{equation}
\frac{\delta{\bar {\cal A}}}{\delta a(x)} = 2\left[\omega - (v_0-c_s)\theta_x\right]a(x)=0,
\end{equation}
which gives us the local dispersion relation, 
\begin{equation}
D_-[x,k=\theta_x;\omega]=0 \qquad \Rightarrow \qquad \omega = [v_0(x)-c_s(x)]\theta_x. 
\end{equation}
Solving this for $\theta_x$, we can now solve for $\theta(x)$
\begin{equation}\label{eq:Phase}
\theta(x) = \theta_0 + \omega \int_{x_0}^x \frac{dx'}{v_0(x')-c_s(x')}.
\end{equation}

Variation of~(\ref{eq:Variation3}) with respect to the phase gives us the action conservation law:
\begin{equation}
\frac{\delta{\bar {\cal A}}}{\delta \theta(x)} = \frac{d}{dx}\left[(v_0-c_s)a^2(x)\right]=0.
\end{equation}
Therefore
\begin{equation}\label{eq:Amplitude}
a^2(x) = \frac{v_0(x_0)-c_s(x_0)}{v_0(x)-c_s(x)}a^2(x_0).
\end{equation}
Clearly, the eikonal approximation breaks down near the event horizon where the denominator goes to zero.

At each $x$, we use the instantaneous polarization by setting $k(x)=\theta_x(x)$ in~(\ref{eq:polarization}), as already mentioned. This, together with the results~(\ref{eq:Phase}) and~(\ref{eq:Amplitude}), give the eikonal solution~(\ref{eq:EikonalAnsatz}).

Therefore, away from the event horizon, the heuristic derivation of the local dispersion function discussed in the first part of Section~\ref{sec:Heuristics} is a useful guide. But, the WKB approximation is invalid when $v_0=\pm c_s$, so how to proceed? We will follow the strategy outlined in Chapter 6 of~\cite{RayTracingBook} by using the insight we have gained from the WKB analysis to construct the normal form, which isolates the tunneling behavior from the other `incoming' wave in the immediate vicinity of the event horizon to the greatest extent possible. 

\section{The Normal Form}\label{sec:NormalForm}

To deal more carefully with the region near the event horizon, we return now to the phase space variational principle~(\ref{eq:Action1}), which we reproduce here for ease of reference. 
\begin{equation}\label{eq:Action5}
{\cal A}'[\psi] = \int dt  \left[\langle {\psi}_t,\psi\rangle_{can}-i\langle {\psi},{\widehat {\cal B}}\psi \rangle_{can} \right].
\end{equation}
Recall the canonical inner product is defined in Eq.~(\ref{eq:CanInner}).

The normal form is developed about a fixed point in ray phase space, $x_0$ and $k_0$. Suppose we have an event horizon where $v_0(0)=c_s(0)$. Then of course, we choose $x_0=0$ as our base point. 

Because the background is time-stationary, we can treat each frequency separately. Choose a fixed, but arbitrary, frequency $\omega$, and introduce the constant (in $x$ and $k$) polarization vectors
\begin{equation}
{\hat {\bf e}}_a\equiv
\frac{1}{\left[2\rho_0c_s k_0(\omega) \right]^{1/2}} 
\left( 
\begin{array}{c}
      1 \\
- i\rho_0c_sk_0(\omega)
\end{array}
\right),
\end{equation}
and
\begin{equation}
{\hat {\bf e}}_b\equiv
\frac{1}{\left[2\rho_0c_s k_0(\omega) \right]^{1/2}} 
\left( 
\begin{array}{c}
      1 \\
 i\rho_0c_sk_0(\omega)
\end{array}
\right).
\end{equation}
The wavenumber $k_0(\omega)$ is chosen so that the non-resonant dispersion relation is satisfied at $x=0$. That is
\begin{equation}
k_0(\omega)\equiv \frac{\omega}{v_0(0)+c_s(0)} = \frac{\omega}{2c_s(0)}.
\end{equation}

Using $k_0(\omega)$ in the polarizations, construct the new ansatz, appropriate for the local region around the event horizon
\begin{equation}\label{eq:NewPsi}
\psi_{\omega}(x,t) = e^{i\left[k_0(\omega)x-\omega t\right]}\left[\phi_a(x){\hat {\bf e}}_a + \phi_b(x){\hat {\bf e}}_b\right].
\end{equation}
Note that we have the identity
\begin{equation}
-i\frac{\partial \psi_{\omega}}{\partial x} = e^{i\left[k_0(\omega)x-\omega t\right]}
\left(k_0(\omega)-i\partial_x \right)\left[\phi_a(x){\hat {\bf e}}_a + \phi_b(x){\hat {\bf e}}_b\right].
\end{equation}
This means we have shifted the origin in ray phase space to $x=0$ and $k=k_0$. Note also that $\psi_{\omega}(x,t)$ is a two-component object, while $\phi_{a,b}(x)$ are scalars. The amplitudes $\phi_{a,b}(x)$ are {\em not} assumed to be eikonal in form. We have made no approximations yet. The wave function~(\ref{eq:NewPsi}) simply reflects a change of dependent variable by choosing a particular polarization basis for a single-frequency solution.

This new form for $\psi_{\omega}$ is inserted into~(\ref{eq:Action5}). Because we have chosen a single frequency for the wave function, we drop the integration over $t$, retaining only the integral over $x$, which gives us the variational principle for the amplitudes:
\begin{eqnarray}\label{eq:Action6}\nonumber
{\cal A}'[\phi_a,\phi_b] &=& \int dx \;\left[\phi_a^*{\widehat D}_{aa}\phi_a +  
                                                    \phi_b^*{\widehat D}_{bb}\phi_b \right. \\
                                   &&  +\left.\phi_a^*{\widehat D}_{ab}\phi_b +  
                                                    \phi_b^*{\widehat D}_{ba}\phi_a \right].
\end{eqnarray}
Here
\begin{equation}
{\widehat D}_{jk} \equiv -\omega {\sf 1}+i{\hat {\bf e}}_j^{\dag}\cdot {\sf J}\cdot 
                    {\widehat {\cal B}}\cdot {\hat {\bf e}}_k, \qquad j,k = a,b.
\end{equation}
Because the polarizations are constant, the operators ${\widehat D}_{jk}$ are clearly just linear combinations of the entries of ${\widehat {\cal B}}$. By construction, the operator-valued matrix $\widehat D$ is self-adjoint with respect to the {\em standard} inner product (recall that ${\widehat {\cal B}}$ is self-adjoint with respect to the {\em canonical} inner product). 

For general background densities and flow profiles, the entries of $\widehat D$ are messy because of the derivatives acting on the background quantities $\rho_0(x)$, $c_s(x)$, and $v_0(x)$, in addition to the action on the amplitudes $\phi_{a,b}$. This obscures what is going on, so let's assume that $\rho(x)=\rho_0=1$ and $c_s(x)=c_s(0)$ are constant, keeping only the variation in $v_0(x)$. (We need the velocity profile to remain a function of $x$ in order to keep the resonance local.)  I.e., we assume that the background variation is on a long spatial scale compared to that of the amplitudes $\phi_{a,b}(x)$. We emphasize that this is a weaker assumption than that used in WKB theory, because we do not assume any special form for the amplitudes $\phi_{a,b}(x)$. 

The calculation described above is most easily done using Weyl symbol methods, as outlined in the discussion leading to~(\ref{eq:NormalForm}). Given the symbol ${\sf D}(x,k;\omega)$, the related operator is given by the usual correspondence 
\begin{equation}
\kappa \quad \leftrightarrow \quad -i\partial_x,
\end{equation}
and the Weyl calculus ensures that we end up with the symmetrized product 
\begin{equation}
x\kappa \quad \leftrightarrow \quad  -\frac{i}{2}\left(x\partial_x+ \partial_x x\right). 
\end{equation}

\section{Summary}

In this paper we have sketched a methodology for studying linear acoustics in transitional regions using ray phase space methods. The main contribution of the paper is to show how to derive the normal form~(\ref{eq:NormalForm}). This isolates the tunneling phenomenon from the non-resonant `incoming' wave, while also providing the leading order coupling between the wave undergoing tunneling and the incoming wave (while the coupling is weak, it is {\em not} zero).

Isolation of the tunneling and casting it into normal form uncovers the coupling constant for that process, and a standard calculation to compute the $\sf S$-matrix connecting the incoming and outgoing tunneling wave fields uncovers the Boltzmann factor in the transmission coefficient, which is at the heart of the theory of Hawking radiation.

In further work, we are pursuing the use of the phase space variational principle~(\ref{eq:Action1}) as a tool for deriving symplectic integrators. Working in ray phase space, as opposed to only $x$- or $k$-space, implies we have a much wider class of transformations (the metaplectic transformations) available to simplify the problem, and to find the best representation for numerical work. We will report on this elsewhere.

\acknowledgements

The authors would like to thank William \& Mary for its support.


\begin{thebibliography}{99}
\bibitem{Tunneling} In this paper, to be consistent with much of the literature on the topic, we use the term {\em tunneling} to refer to a process whereby one incoming ray can `split' into two outgoing rays, but that only a single polarization is involved, hence the problem can locally be reduced to a scalar one by Galerkin methods. If two polarizations are involved in the resonance, then we can only reduce the problem to a local $2\times 2$ representation, which is called {\em mode conversion}. Hence, using this terminology, the Hawking process is a kind of tunneling process, not a mode conversion. The presence of the `incoming' wave in the midst of the tunneling region means the problem is not standard, however.
\end{thebibliography}
\end{document}